\documentclass[twocolumn,amsmath,amssymb,prb,longbibliography]{revtex4-1}
\usepackage{graphicx}
\usepackage{dcolumn}
\usepackage{bm}
\usepackage{comment}
\usepackage{color}
\DeclareMathAlphabet \mathbfcal{OMS}{cmsy}{b}{n}
\begin{document}
	
\title{
Fundamentally fastest optical processes at the surface of a topological insulator
}

\author{S. Azar Oliaei Motlagh}
\author{Jhih-Sheng Wu}
\author{Vadym Apalkov}
\author{Mark I. Stockman}
\affiliation{Center for Nano-Optics (CeNO) and
Department of Physics and Astronomy, Georgia State
University, Atlanta, Georgia 30303, USA
}

\date{\today}
\begin{abstract}
We predict that a single oscillation of a strong optical pulse can significantly populate the surface conduction band of a three-dimensional topological insulator, $\mathrm {Bi_2Se_3}$. Both linearly- and circularly-polarized pulses generate chiral textures of interference fringes of population in the surface Brillouin zone. These fringes constitute a self-referenced electron hologram carrying information on the topology of the surface Bloch bands, in particular, on the effect of the warping term of the low-energy Hamiltonian. These electron-interference phenomena are in a sharp contrast to graphene where there are no chiral textures for a linearly-polarized pulse and no interference fringes for circularly-polarized pulse. These predicted reciprocal space electron-population textures can be measured experimentally by time resolved angle resolved photoelectron spectroscopy (TR-ARPES) to gain direct access to non-Abelian Berry curvature at topological insulator surfaces. 
%
\end{abstract}

\maketitle

\section{Introduction} 
\label{intro}

Topological insulators (TI's) represent a modern class of crystalline materials where the bulk is semiconducting and the surfaces are semimetallic \cite{Mele_et_al_PhysRevLett.98_2007_Topological_Insulators, Hsieh_et_al_Science_2009_Quantum_Spin_Textures_in_TIs, Hassan_Kane_Topological_insulators_RevModPhys_2010, Bansil_et_al_RevModPhys.88_2016_Topological_Band_Theory}. The $\Gamma$-point of the surface-state band is a Dirac point where the electron dispersion is linear as characteristic of Dirac fermions. There is locking of spin and linear momentum caused by spin-orbit interaction. This provides protection against back scattering and Anderson localization.  The absence of the bandgap and, consequently, the linear electron dispersion at the $\Gamma$-point, are protected by time-reversal ($\cal T$) symmetry. The Bloch bands near the Dirac point are chiral and carry the Berry phase of $\pm\pi$.

To resolve chirality of the surface states of TI's and related spin textures one has to use chiral tools: circularly polarized excitation radiation \cite{Gedik_et_al_PhysRevLett.107_2011_Warped_Helical_Spin_Textures_in_TI, Gedik_et_al_Science_2013_Floquet_Bloch_States_In_Topological_Insulators, Rader_et_al_PhysRevX.4_2014_Photoemission_of_Topological_Insulator} or detection of the spin state of electrons in spin resolved angle resolved photoemission spectroscopy (SR-ARPES) \cite{Hsieh_et_al_Science_2009_Quantum_Spin_Textures_in_TIs}. A unique property of the surface bands of a TI -- band-dependent locking of the spin and momentum -- allows one to manipulate the spin states by controlling the electron momentum in the Bloch bands. Keeping in mind both fundamental interest and spintronics applications, it is important to perform such manipulation as fast as possible, that is within a single optical cycle.  Such ultrafast manipulation would allow avoiding significant relaxation, including quantum decoherence during the excitation cycle. It also would open up unique possibilities for ultrafast, PHz-band scale, information processing. Note that the existing experiments on TI's employed relatively long ($\gtrsim 100$ fs) excitation pulses of moderate field amplitudes \cite{Gedik_et_al_Science_2013_Floquet_Bloch_States_In_Topological_Insulators}, $F_0\lesssim 10^{-3}~\mathrm{V/\AA}$.

The goal of this article is to show theoretically for a 3D TI a possibility to manipulate electron population, crystal momentum and, consequently spin, in a fundamentally fastest way -- during only single cycle of a chiral optical pulse with a moderately strong optical field, $F_0\sim 0.05-0.2~\mathrm{V/\AA}$. We have predicted such ultrafast processes earlier for graphene both for linear polarization \cite{Stockman_et_al_PRB_2016_Graphene_in_Ultrafast_Field} and circular polarization \cite{Stockman_et_al_PhysRevB.93.155434_Graphene_Circular_Interferometry, Stockman_et_al_PhysRevB.96_2017_Berry_Phase}. Electrical currents and charge transfer associated with ultrafast strong-field excitation of graphene have been recently observed experimentally \cite{Higuchi_Hommelhoff_et_al_Nature_2017_Currents_in_Graphene}. The underlying chiral distributions of excited electrons in the full Brillouin zone in both the conduction band (CB) and valence band (VB) can fundamentally be observed using time resolved ARPES (TR-ARPES) \cite{Chiang_et_al_PhysRevLett.107_Berry_Phase_in_Graphene_ARPES, Gierz_Snapshots-non-equilibrium-Dirac_Nat-Material_2013, Crepaldi_et_al_PhysRevB.88.121404_2013_Electron_Phonon_Scatering_in_BiSe}. However, such studies have not yet been done.

We predict that, similar to graphene, a significant CB population is induced during a single optical oscillation. The main mechanism responsible for this process is Bloch motion of electrons in reciprocal space induced by strong optical fields. In contrast to graphene, the resulting electron CB distribution in the reciprocal space is highly anisotropic and chiral; to a significant degree it is defined by the so-called warping terms in the effective Hamiltonian \cite{Liu_PRB_2010_Model_Hamiltonian_3D_TI}. The electron relaxation, which follows the excitation pulse, will cause surface currents and generation of THz radiation. The latter also provides an access to the initial CB electron distribution \cite{Braun-Nat_Commun_2016-Ultrafast-photocurrents-at-the-surface-of-TI}. Yet another approach to get access to the electron distribution of the surface CB in TIs is ultrafast time-resolved transient reflectivity and Kerr rotation \cite{Qi_et_al_PhysRevLett.116.036601_Photoinduced_Spin_Dynamics_in_TI_BiSe}.




Specifically, in this work, we study ultrafast electron dynamics on the surfaces of a 3D TI $\mathrm {Bi_2Se_3}$, which has a bulk bandgap of $\approx$ 0.3 eV and gapless surface states \cite{Liu_et_al_PRB_2010_Model_Hamiltonian, Cava_et_al_2013_Crystal_Bi2Se3, Glinka_et_al_PRB_2015_TI_Crystal}. These gapless surface states are protected by the time reversal symmetry and possess a single Dirac cone located at the $\Gamma$ point in the reciprocal space \cite{Liu_et_al_PRB_2010_Model_Hamiltonian, Zhang_et_al_nature_2009_TI,
Zhang_et_al_Nature_Physics_2009_Topological_Insulator_Bi2Se3, Feng_et_al_2012_TI}. Similar to graphene, the interband dipole coupling at the surface of 3D TI is singular at the Dirac point, and the energy dispersion near the Dirac point is linear (quasi-relativistic). The main difference from graphene is that the effective low-energy Hamiltonian in 3D TI's has large nonlinear (cubic) terms. Such terms, which are known as the warping terms, lower the point symmetry of the system to three-fold (C$_3$) \cite{Liu_PRB_2010_Model_Hamiltonian_3D_TI}. This lower symmetry results in unique features in the ultrafast electron dynamics. In particular, as we show below in this article, for a single-oscillation circularly-polarized pulse, the CB population distribution in the reciprocal space shows a pronounced chirality and a pattern of interference fringes.


Because surface Bloch bands of a TI are gapless, the processes of the electron transfer between the valence band (VB) and CB is non-resonant and, consequently, broadband. Therefore, the dependence on the mean (carrier) frequency or duration of the pulse is very smooth. As an example we choose a pulse with duration of 5 fs (carrier frequency $\bar\omega\approx 1.2~\mathrm{eV/\hbar}$). This duration is chosen to be shorter than the fastest electron scattering time \cite{Braun-Nat_Commun_2016-Ultrafast-photocurrents-at-the-surface-of-TI} in TI's, which is $\gtrsim 10$ fs. The pulse electric field causes electron motion in the reciprocal space within each band where the crystal momentum excursion is defined by the Bloch acceleration theorem -- see Eq.\ (\ref{kvst}) below. 
The $\mathrm{VB\to CB}$ transitions occur when an electron passes in a vicinity of the Dirac point ($\Gamma$ point) or other points where the interband transition dipole matrix element is enhanced. The excursion of electron crystal momentum during the pulse can estimated as $\Delta k\sim \pi e F_0/(\hbar\bar\omega)$, where $e$ is unit charge and $F_0$ is the field amplitude. To travel a half of the Brillouin zone, this excursion should be $\Delta k\sim \pi/a$, where $a$ is a lattice constant. Assuming $a\approx 5~\mathrm\AA$, we obtain $F_0\approx \hbar\bar\omega/(ea)\approx 0.2~\mathrm{V/\AA}$. Correspondingly, we consider a field amplitude range that includes this value, $F_0=0.05-0.5~\mathrm{V/\AA}$.

The paper is organized as follows. In Section II, a model and main equations, which are used to calculate the electron dynamics in the presence of ultrafast external electric fields, are introduced. In Section III, the results for linearly and circularly polarized pulses are presented and discussed. 

\section{MODEL AND MAIN EQUATIONS}

\begin{figure}
\begin{center}\includegraphics[width=0.47\textwidth]{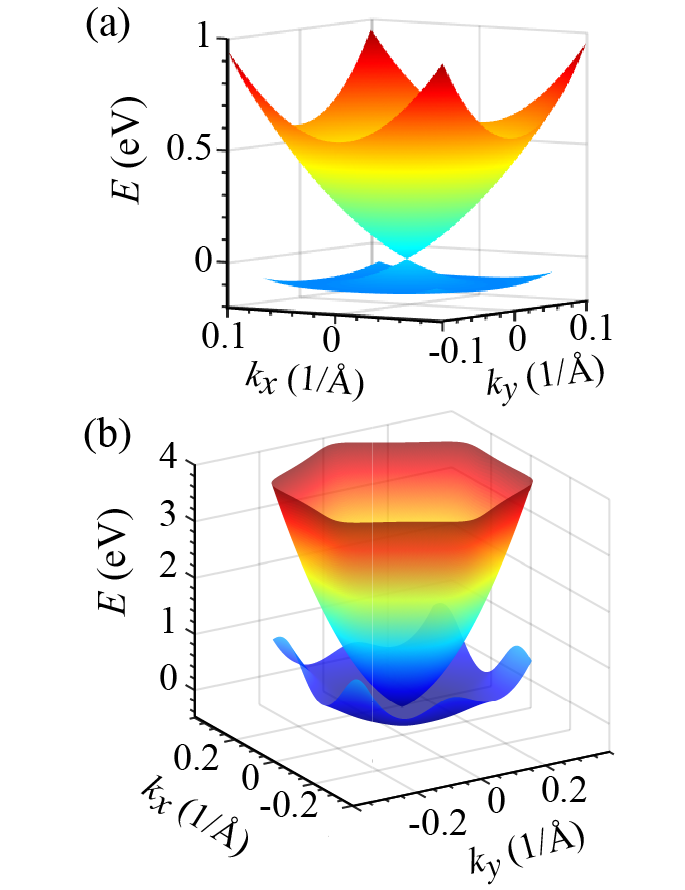}\end{center}
\caption{(a) Low-energy electron dispersion of the surface state around the Dirac point, i.e.,   the $\mathrm{\Gamma}$-point at $\mathbf k=(0,0))$,  as a function of wave vector $\mathbf k$. The Dirac cone at the Gamma point is clearly visible. (b) Energy dispersion in a wide range where the sixth-order warping is manifest.}
  \label{fig:Energy_TI}
\end{figure}

An effective low-energy surface Hamiltonian of $\mathrm {Bi_2Se_3}$ near the Dirac point has the following form  \cite{Liu_PRB_2010_Model_Hamiltonian_3D_TI}
\begin{equation}
H_0=A_1{k^2}+A_2(\sigma_xk_y-\sigma_yk_x)+A_3(k^3_++k^3_-)\sigma_z ,
\label{H0}
\end{equation}
where $\sigma_x$, $\sigma_y$, and $\sigma_z$ are Pauli matrices, $(k_x,k_y)$ is crystal momentum, $k_{\pm}=k_x \pm ik_y=k e^{\pm i \theta}$, $k=\left(k_x^2+k_y^2\right)^{1/2}$, and $A_1$, $A_2$,  and $A_3$  are constants  that  are equal to  23.725 $ \mathrm{eV\AA^2 }$, 3.297 $\mathrm{eV\AA}$, and 25.045 $ \mathrm{eV\AA^3}$, respectively. The cubic term in this low-energy Hamiltonian is called the hexagonal warping term \cite{Fu_PRL_2009_Warping}.
The energies of the VB and the CB can be found from the above Hamiltonian as
\begin{eqnarray}
E_{c}(\mathbf k)&=&A_1 k^2+\sqrt{A^2_2 k^2+ 4A^2_3k^2_x(k^2_x-3k^2_y)^2}~~,
\nonumber \\
E_{v}(\mathbf k)&=&A_1 k^2 -\sqrt{A^2_2 k^2+ 4A^2_3k^2_x(k^2_x-3k^2_y)^2}~~,
\label{Energy_TI}
\end{eqnarray}
where indices $c$ and $v$ stand for the CB and VB, respectively. Near the $\Gamma$-point (for $k\to0$), this dispersion simplifies to two Dirac cones: $E_{c,v}=\pm A_2 k$. This energy dispersion is displayed in Fig. \ref{fig:Energy_TI} where the $\mathrm{\Gamma}$ point is at $\mathbf k=(0,0)$. Below we assume that the system is undoped, with the Fermi energy at 0, where the VB is fully occupied and the CB is completely empty.
This energy dispersion has sixfold 
symmetry (see Fig. \ref{fig:Energy_TI}(b)), which is due to the warping term. 

As we have already argued above in Sec.\ \ref{intro}, during an ultrashort optical pulse with the duration of $\sim5$ fs, electron dynamics is coherent since the electron scattering times in $\mathrm{Bi_2Se_3}$ are $\gtrsim20$ fs. Then the evolution of the system can be described by the time-dependent Schr\"odinger equation (TDSE),
\begin{equation}
i\hbar \frac{{d\Psi }}{{dt}} = { H(t)} \Psi , 
\label{Sch}
\end{equation}
with Hamiltonian
\begin{equation}
{ H}(t) = { H}_0 - e{\bf{F}}(t){\bf{r}},
\label{Ht}
\end{equation}  
where $\mathbf F(t)$ is the pulse's electric field, and $e$ is electron charge.

We will be employing moderately high field amplitudes, $F_0\gtrsim0.1~\mathrm{V/\AA}$. At such intensities, the number of photons, $N_p$, per pulse within the minimum coherence area of $\sim\lambda^2$, where $\lambda\sim 1~\mathrm{\mu m}$ is wavelength,  
\begin{equation}
N_p\sim \frac{c \tau_p \lambda^2 F_0^2}{4\pi \hbar\bar\omega}\sim 5\times 10^7~,
\end{equation}
where $c$ is speed of light; we assume realistic parameters: $\tau_p\sim3~\mathrm{fs}$ is the pulse duration, and $\hbar\bar\omega\sim 1~\mathrm{eV}$ is the mean photon energy. With such a large photon number involved, it is legitimate to describe $\mathbf F(t)$ as a classical electric field keeping quantum-mechanical description of the solid. This is a usual approach in high-field optics -- see, e.g., Refs.\ \onlinecite{Corkum_Krausz_Nature_Physics_2007_Attosecond_Science, Krausz_Ivanov_RevModPhys.81.163_2009_Attosecond_Review, Krausz_Stockman_Nature_Photonics_2014_Attosecond_Review}. Note that quantized optical fields are used for much lower intensities 
\cite{Kibis_PhysRevB.81_2010_MI_Transition_in_Graphene_by_CP_Photons, Kristinsson_Sci_Rep-2016-Control_of_electronic_transport_in_graphene}; such an approach is not needed at our fields. With classical field $\mathbf F(t)$, we solve the Schr\"odinger equation in the basis chosen numerically without further approximations. Our pulse is just a single optical oscillation; therefore field $\mathbf F(t)$ is not periodic, and its effect cannot be described as band gap opening as in Ref.\ \onlinecite{Kibis_PhysRevB.81_2010_MI_Transition_in_Graphene_by_CP_Photons}. However, the dynamic Stark effect and other field-dressing effects during the pulse are fully taken into account by our solution.

In solids, the applied electric field generates both the intraband and interband electron dynamics. The intraband dynamics is determined by the Bloch acceleration theorem \cite{Bloch_Z_Phys_1929_Functions_Oscillations_in_Crystals}, 
\begin{equation}
\hbar \frac{{d{\bf{k}}}}{{dt}} = e{\bf{F}}(t).
\label{acceleration}
\end{equation}
From this, for an electron with an initial crystal momentum ${\bf q}$,   time-dependent crystal momentum ${\mathbf k }({\mathbf q},t )$ is expressed as
\begin{equation}
{{\bf{k}}}({\bf{q}},t) = {\bf{q}} + \frac{e}{\hbar }\int_{ - \infty }^t {{\bf{F}}({t^\prime})d{t^\prime}}. 
\label{kvst}
\end{equation}
The corresponding wave functions, which are solutions of the Schr\"odinger equation (\ref{Sch}) within a 
single band $\alpha$, i.e., without interband coupling, are the well-known Houston functions \cite{Houston_PR_1940_Electron_Acceleration_in_Lattice},
\begin{equation}
\Phi _{\alpha {\bf{q}}}^{(H)}({\bf{r}},t) = \Psi _{{{\bf{k}}}({\bf{q}},t)}^{(\alpha )}({\bf{r}}){e^{ - \frac{i}{\hbar }\int_{ - \infty }^t {d{t_1}{E_\alpha }[{{\bf{k}}}({\bf{q}},{t_1})]} }}~,
\end{equation}
where $\alpha=v,c$ for the VB and CB, correspondingly, and $ \mathrm{\Psi^{(\alpha)}_{{\mathbf k}}} $ are Bloch-band eigenfunctions in the absence of the pulse field, and $E_\alpha(\mathbf k)$ are the corresponding eigenenergies.

The interband electron dynamics is determined by the solution of the TDSE (\ref{Sch}).  Such a solution can be expressed in the basis of the Houston functions $\Phi^{(H)}_{\alpha {\bf q}}({\bf r},t)$,
\begin{equation}
\Psi_{\bf q} ({\bf r},t)=\sum_{\alpha=c,v}\beta_{\alpha{\bf q}}(t) \Phi^{(H)}_{\alpha {\bf q}}({\bf r},t),
\end{equation}
where 
$\beta_{\alpha{\bf q}}(t)$ are expansion coefficients, which satisfy the following system of coupled equations
\begin{eqnarray}
&&\frac{{d{\beta _{c{\mathbf{q}}}}(t)}}{{dt}} =  - \frac{i}{\hbar }\mathbf F(t) \mathbf Q_{cv}(\mathbf q, t) \beta _{v \mathbf{q}}(t) ,
\nonumber \\
&&\frac{{d{\beta _{v{\bf{q}}}}(t)}}{{dt}} =  - \frac{i}{\hbar } \mathbf{F}(t)\mathbf{Q}^\ast_{cv}(\mathbf q,t)\beta _{c\mathbf{q}}(t) ,
\label{eq:beta_1,2}
\end{eqnarray}
where
\begin{eqnarray}
&&\mathbf Q_{cv}(\mathbf q,t)=
\mathbf D_{cv}[\mathbf k (\mathbf q,t)]\exp\left(i\phi^\mathrm{(d)}_{cv}(\mathbf q,t)\right),
 \label{Q}
\\
&&\phi^\mathrm{(d)}_{cv}(\mathbf q,t)=
\nonumber
\\
 &&- \frac{1}{\hbar} \int_{-\infty}^t dt^\prime \left(E_c[\mathbf k (\mathbf q,t^\prime)]-E_{v}[\mathbf k (\mathbf q,t^\prime)]\right),
 \label{phi}
 \\ 
&&\mathbf D_{cv}(\mathbf q)=e \mathbfcal{A}_{cv}(\mathbf q); ~
{\mathbfcal{A}}_{cv}({\mathbf q})=
\left\langle \Psi^{(c)}_\mathbf q  |   i\frac{\partial}{\partial\mathbf q}|\Psi^{(v)}_\mathbf q   \right\rangle .
\label{D}
\end{eqnarray} 
Here $ \mathrm{\Psi^{(v)}_{{\mathbf q}}} $ and $ \mathrm{\Psi^{(c)}_{{\mathbf q}}} $ are eigenfunctions of the Hamiltonian without an optical field, $H_0$;   matrix element ${\mathbfcal A}_{cv}(\mathbf q)$ is the well-known non-Abelian Berry connection \cite{Wiczek_Zee_PhysRevLett.52_1984_Nonabelian_Berry_Phase, Xiao_Niu_RevModPhys.82_2010_Berry_Phase_in_Electronic_Properties, Yang_Liu_PhysRevB.90_2014_Non-Abelian_Berry_Curvature_and_Nonlinear_Optics}; $\mathbf D_{cv}(\mathbf q)$ is the interband dipole matrix element, which determines optical transitions between the VB and the CB at crystal momentum $\mathbf q$, and $\phi^{\mathrm{(d)}}_{cv}(\mathbf q,t)$ is the dynamic phase; the trajectory in the reciprocal space, $\mathbf k(\mathbf q, t)$, is given by the Bloch theorem (\ref{kvst}). 

Note that fundamentally Eq.\ (\ref{eq:beta_1,2}) is a Schr\"odinger equation in the interaction representation in the adiabatic basis of the Houston functions, where the wave function is a two-component state vector $\left(\beta_{c\mathbf q},\beta_{v\mathbf q}\right)$. This equation defines the exact solution for dynamics of the system, which is limited only by the size of the basis set (i.e., truncation of the Hilbert space of the surface states of the TI). In particular, it contains such phenomenon as bandgap opening in the field of a circularly-polarized pulse

We introduce components of non-Abelian Berry connection as a vector $\left\{\mathcal A_x(\mathbf k), \mathcal A_y(\mathbf k)\right\}=\mathbfcal A_{cv}(\mathbf k)$. Substituting the eigenfunctions of the field-free Hamiltonian (\ref{H0}) into Eq.\ (\ref{D}), we obtain for its components
\begin{eqnarray}
\mathcal A_x(\mathbf k)&=&\mathcal N\Bigg(-\frac{1}{2} \frac{k_y}{k^2}
\nonumber\\
&&-i \frac{2k^4_x+3k^2_y(k^2_x-k^2_y)}{k^2 \sqrt {4k^2_x(k^2_x-3k^2_y)^2+ (\frac{A_2}{A_3}k)^2}}\Bigg)~,
\label{Dx_TI}
\\
\mathcal A_y(\mathbf k)&=&\mathcal N\Bigg(\frac{1}{2} \frac{k_x}{k^2}
\nonumber\\
&&
 +i \frac{k_xk_y(7k^2_x+3k^2_y)}{k^2 \sqrt {4k^2_x(k^2_x-3k^2_y)^2+ (\frac{A_2}{A_3}k)^2}}\Bigg)~,
 \label{Dy_TI}
\end{eqnarray}
\begin{figure}
\begin{center}\includegraphics[width=0.47\textwidth]{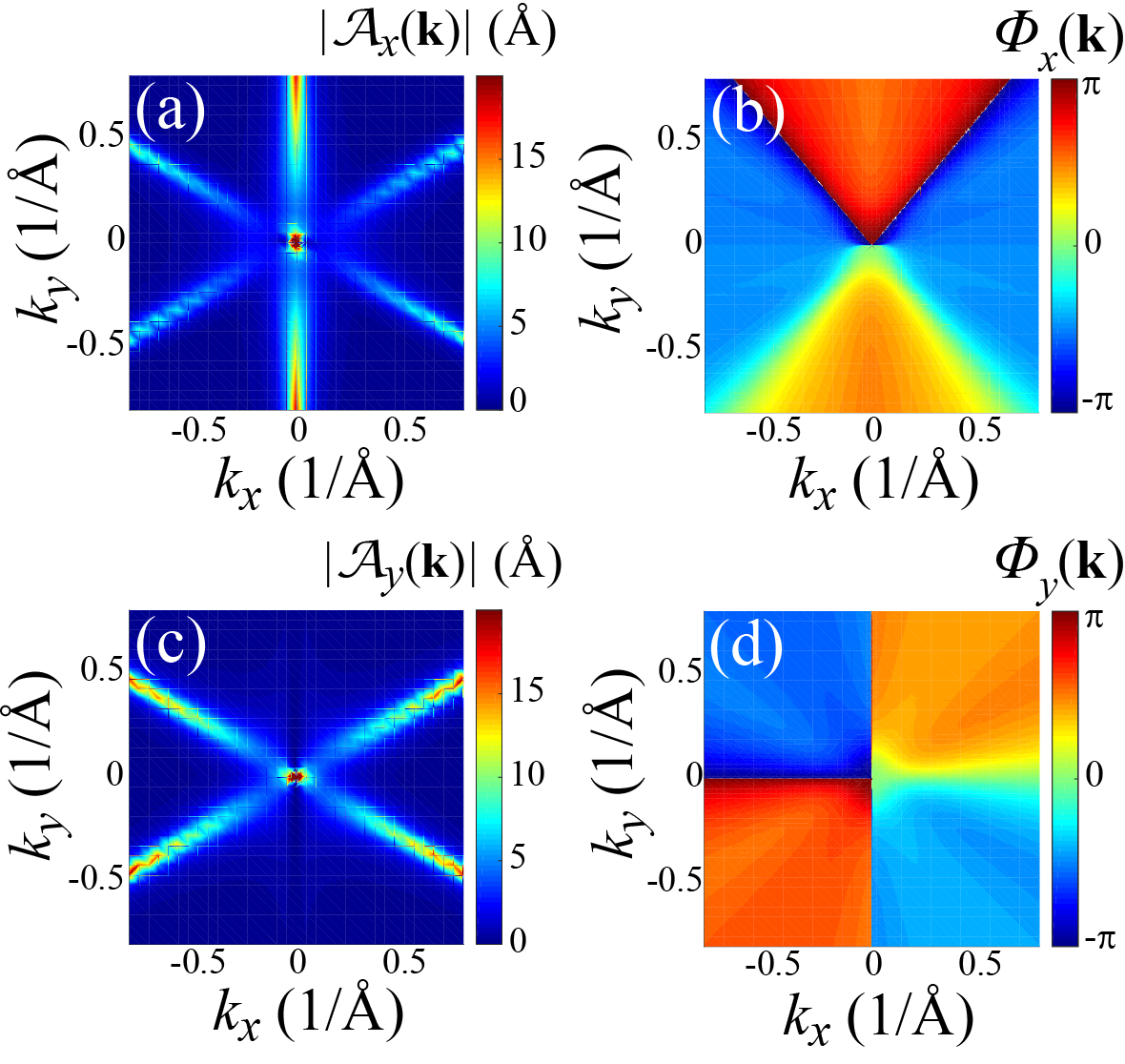}\end{center}
  \caption{Complex vector of non-Abelian Berry connection $\mathbfcal A=\mathbfcal A_{cv}$.
  (a) Magnitude of $x$-component $\left|\mathcal A_x(\mathbf k)\right|$ (in units of $\mathrm{\AA}$) as a function of crystal momentum $\bf k$. (b)  Phase  $\Phi_x(\mathbf k)=\arg\left(\mathcal A_x(\mathbf k)\right)$ as a function of crystal momentum $\mathbf k$. (c) Same as (a) but for $\mathcal A_y(\mathbf k)$. (d) Same as (b) but for $\mathcal A_y(\mathbf k)$.}
  \label{fig:Dipole}
\end{figure}%
where normalization coefficient is
\begin{equation}
\mathcal N=\left(1+ \frac{4k^2_x(k^2_x-3k^2_y)^2}{(\frac{A_2}{    A_3}k)^2} \right)^{- \frac{1}{2}}.
\nonumber
\end{equation}
 Similar to graphene, there are singularities at the Dirac point, $\mathbf k=0$, for both $\mathcal A_x({\bf k})$ and $\mathcal A_y({\bf k})$  -- see Fig. \ref{fig:Dipole}. However, unlike graphene, $\mathcal A_x({\bf k})$ and $\mathcal A_y({\bf k})$ 
have both imaginary and real parts due to the warping term in surface Hamiltonian (\ref{H0}).
Consequently, there is a nontrivial dependence on angle, $\tan^{-1}(k_y/k_x)$:
the components of the non-Abelian Berry connection, $\mathcal A_x$ and $\mathcal A_y$, exhibit sharp maxima along six and four ridges, which are clearly seen in Figs.\  \ref{fig:Dipole} (a) and (c), respectively. 

When an electron is compelled by the optical field to move within a band, the transitions will occur when its trajectory ${{\bf{k}}}({\bf{q}},t)$, which is defined by the Bloch acceleration theorem  (\ref{kvst})], crosses these ridges and passes in the vicinity of the Dirac point. The corresponding transition amplitudes will interfere generating a texture of fringes in the reciprocal space.


We solve Eqs. (\ref{eq:beta_1,2}) numerically with an initial condition corresponding to an occupied VB and an empty CB ($\beta _{v {\bf q}} = 1$ and $\beta _{c {\bf q}} = 0$). Below we characterize the electron dynamics in terms of the residual CB population as a function of crystal momentum, $N_\mathrm{CB}^{(\mathrm{res})}(\mathbf{q})=|\beta _{{c}{\mathbf{q}}}(t=\infty)|^2$, and the mean residual CB population 
\begin{equation}
n_{c} = \left\langle N_\mathrm{CB}^{(\mathrm{res})}(\mathbf{q})\right\rangle_\mathbf q~,
\end{equation}
where $\left\langle \dots\right\rangle_\mathbf q$ denotes average over crystal momentum $\mathbf q$ in the first Brillouin zone.
 
Compare results for 3D TI with the corresponding results for graphene where low-energy Dirac Hamiltonian is
\begin{equation}
H_{0}^\mathrm{(gr)}=\hbar v_F (\sigma_x k_x + \sigma_yk_y),
\label{H0gr}
\end{equation}
where $v_F\sim10^6~\mathrm{ m/s}$ is the Fermi velocity. Then the energies of the CB and VB are 
\begin{eqnarray}
E_{c}^\mathrm{(gr)}(\mathbf k)&=& \hbar v_F k \nonumber \\
E_{v}^\mathrm{(gr)}(\mathbf k)&=& - \hbar v_F k~.
\label{Energy_TI_gr}
\end{eqnarray}
The corresponding components of the non-Abelian Berry connection are  
\begin{eqnarray}
\mathcal A_x^\mathrm{(gr)}(\mathbf k)&=&  \frac{-\frac{1}{2}k_y}{k_x^2+k_y^2} ~,
\label{Axgr}\\
\mathcal A_y^\mathrm{(gr)}(\mathbf k)&=&  \frac{\frac{1}{2}k_x}{k_x^2+k_y^2}~.
\label{Aygr}
\end{eqnarray}
These are obviously real quantities in contrast to Eqs.\ (\ref{Dx_TI}) and (\ref{Dy_TI}).

\section{Results and discussion}
\subsection{Linearly polarized pulse}
\label{Linear_Polarized}

\begin{figure}
\begin{center}\includegraphics[width=0.49\textwidth]{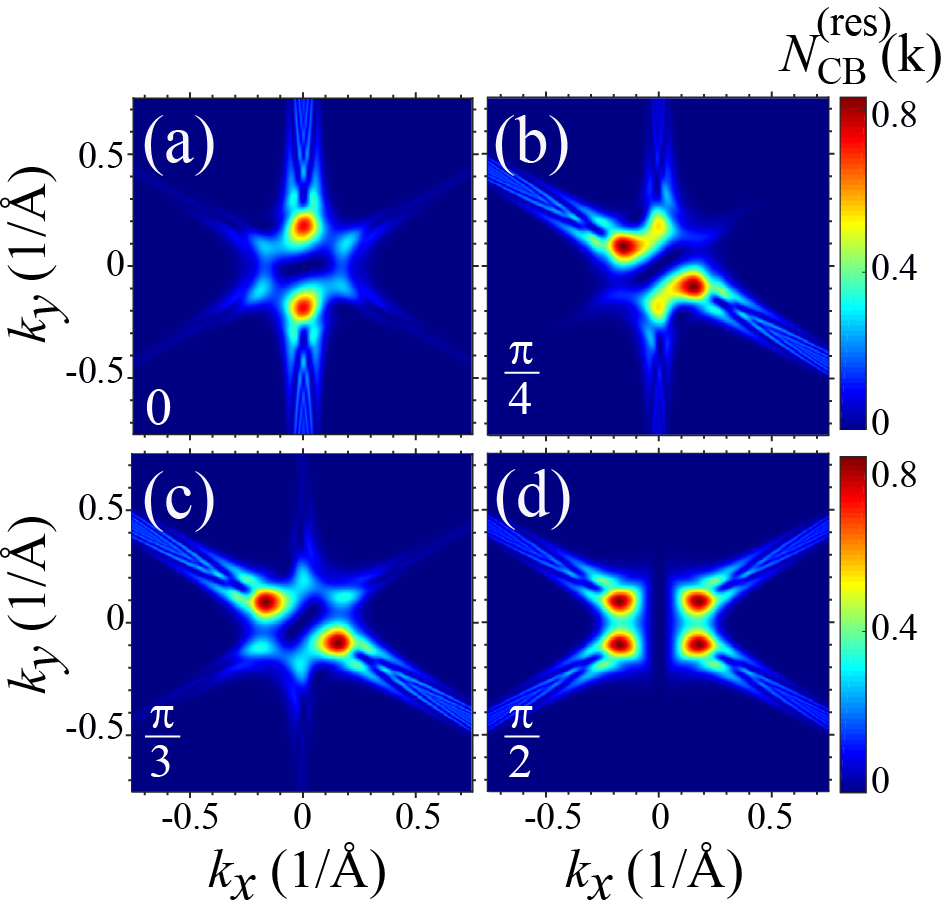}\end{center}
  \caption{(Color online) Residual CB population, $N^{\mathrm{(res)}}$, on the surface of  $\mathrm{Bi_2Se_3}$ as a function of wave vector $\bf k$ for different values of angle $\theta$. The amplitude of the
  electric field is  0.10 $\mathrm{V/\AA}$. The angle is (a) $\theta= 0$, (b) $\theta= \pi/4$, (c) $\theta= \pi/3$, and (d) $\theta= \pi/2$. }
  \label{fig:RCB_TI}
\end{figure}

We apply a linearly-polarized single-oscillation optical pulse incident normally to the surface of 3D TI.  The electric field waveform of the pulse is parametrized in the following form
\begin{equation}
\mathbf F(t)=\left\{F_x(t),F_y(t) \right\}=F(t)\left\{\cos\left(\theta\right),\sin\left(\theta\right\} \right) 
\end{equation}
where $ F(t)=F_0e^{-u^2}(1-2u^2)$, $F_0$ is the amplitude of the field, $u=t/\tau$, $\tau=1$ fs, and $\theta$ is the polarization angle of the applied field.
Before the pulse, the CB is empty, and the VB is fully occupied. During the pulse, the electron dynamics is characterized by redistribution of electrons between the VB and the CB. 
After the pulse, there is a significant residual CB population -- see Fig. \ref{fig:RCB_TI} -- manifesting irreversibility of the electron due to related to the gapless spectrum of the TI's surface states.  Note that an earlier work \cite{Schiffrin_at_al_Nature_2012_Current_in_Dielectric, Schiffrin_at_al_Nature_2012_Current_in_Dielectric, Stockman_et_al_Sci_Rep_2016_Semimetallization, Krausz_Stockman_Nature_Photonics_2014_Attosecond_Review} showed that strong-field induced CB population is highly reversible, i.e., disappearing after the pulse end, for insulators such as silica and alumina (quartz and sapphire). This reversibility is due to the presence of a wind bandgap ($\approx 10$ eV) between the VB and the CB, which significantly exceeds the characteristic frequency of the excitation pulse. Such a condition causes the CB population to adiabatically follow the magnitude (modulus) of the pulse field, $\left| F(t)\right|$. This is not the case for TI's since their surface-band spectrum is gapless.

The distribution of the residual CB population in the first Brillouin zone of the surface bands of $\mathrm{Bi_2Se_3}$ is shown in Fig. \ref{fig:RCB_TI} for different angles between the polarization direction and wthe positive $x$ axis:  (a) $\theta= 0$, (b) $\theta= \pi/4$, (c) $\theta= \pi/3$, and (d) $\theta= \pi/2$. The amplitude of the electric field is $F_0=0.1~\mathrm{V/\AA}$. There is a pattern of ``hot spots'' with a large CB population and interference fringes are clearly visible. The pair of hot spots seen in each panel is an image of the Dirac point split by a dark line passing through the $\Gamma$-point in the polarization direction. Note that this dark line originate from electron passing twice through a Dirac point, which results in its return back to the VB and the zero CB population. This can also be interreted as a result of the pseudospin conservation. The interference fringes farther from the Dirac popint in Figs.\ \ref{fig:RCB_TI} (a)-(c) show a pronounced chirality. These fringes originate from the electron $\mathrm{VB\to CB}$ transitions caused the warping term and replicate its symmetry. 

The components of the non-Abelian Berry connection, which are responsible for the interband coupling for a given direction of polarization, are 
$\mathcal A_x(\mathbf k)$ and $\mathcal A_y(\mathbf k)$  for $\theta=0$ and $\theta=\pi/2$, respectively.  The distribution of $N_{\mathrm{CB}}^{\mathrm{(res)}}(\mathbf k)$ follows the profile of the corresponding components, $\mathcal A_x(\mathbf k)$ and $\mathcal A_y(\mathbf k)$, see   Fig. \ref{fig:RCB_TI} (a) and (d), respectively.


\begin{figure}
\begin{center}\includegraphics[width=0.47\textwidth]{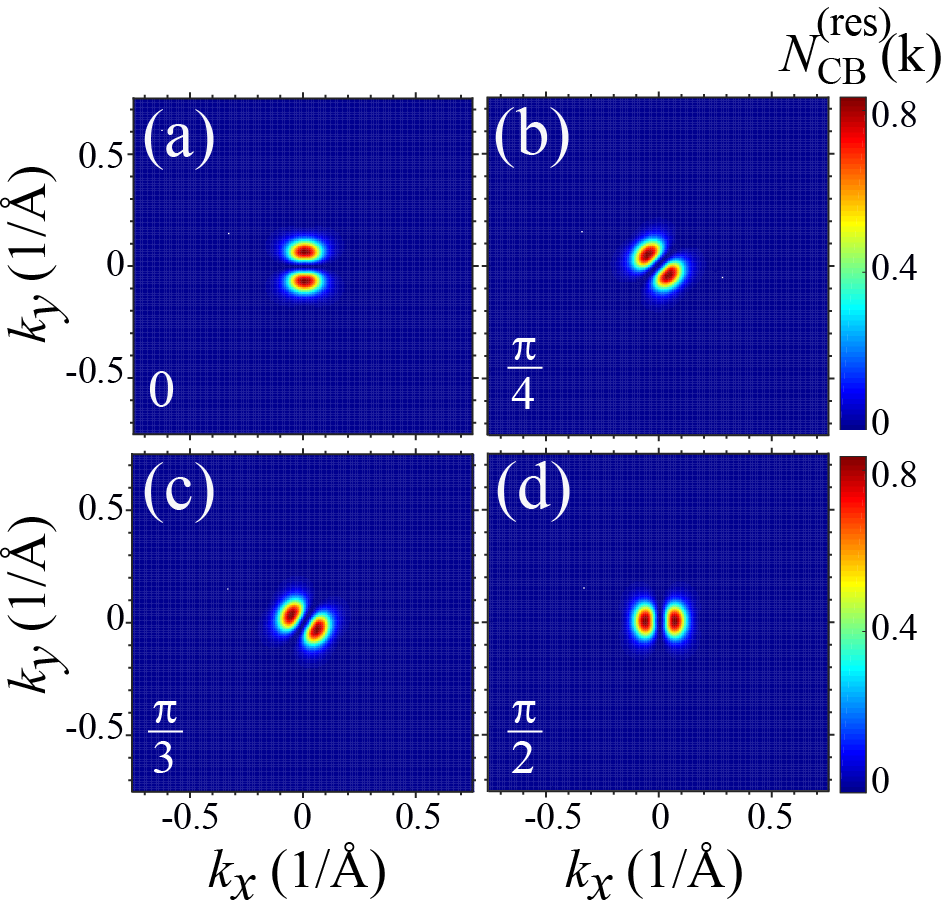}\end{center}
\caption{(Color online) Residual CB population, $N^{\mathrm{(res)}}_{\mathrm{CB}}$, of graphene as a function of wave vector $\bf k$ for different values of angle $\theta$. The amplitude of the electric field is 0.10 $\mathrm{V/\AA}$. The angle is (a) $\theta= 0$, (b) $\theta= \pi/4$, (c) $\theta= \pi/3$, and (d) $\theta= \pi/2$. The dynamics of the graphene system is modeled within the low energy effective Dirac model.   }
\label{fig:RCB_graphene}
\end{figure}

\begin{figure}
\begin{center}\includegraphics[width=0.45\textwidth]{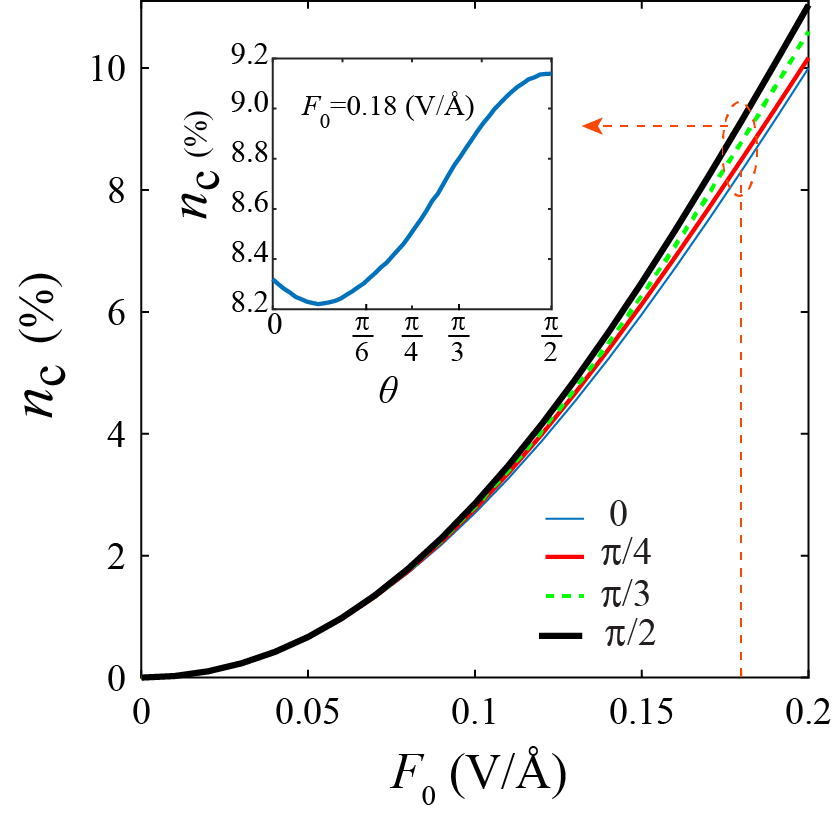}\end{center}
\caption{(Color online) Mean residual CB population $n_c$ as a function of field amplitude $ F_0$ for different values of $\theta$: $\theta= 0$ (blue solid line), $\theta= \pi/4$ (red solid line), $\theta= \pi/3$
(green dashed line), and $\theta= \pi/2$ (black solid line). Inset: total residual CB population, $n_c$, as a function of $\theta$ for the amplitude of the electric field of 
 0.18 $\mathrm {V/\AA}$. }
  \label{fig:Edited_N}
\end{figure}

This behavior of the residual CB population is similar but with significant differences to what is calculated for graphene near the Dirac points (K or K$^\prime $), 
see Fig.\ \ref{fig:RCB_graphene}, where the residual CB population is shown for a low-energy effective model of graphene [Eq.\ (\ref{H0gr})]  for 
different polarization directions. The distribution of $N^{\mathrm{(res)}}_\mathrm{CB}(k)$ for graphene is symmetric with respect to the polarization plane in contrast to the 3D TI, cf.\ Fig.  \ref{fig:RCB_TI}, where this symmetry is broken by the warping terms.

There is also another fundamental physical difference between the electron-distribution textures in graphene and TI. In graphene, there is locking of crystal momentum and pseudospin, and this locking is opposite in the VB and CB. For TIs, there is locking of crystal momentum and real electron spin. Thus the textures of the electron population in the reciprocal space are simultaneously spin-polarization textures. Potentially, such momentum/spin textures can be directly measured using spin-polarized TR-ARPES \cite{Hsieh_et_al_Science_2009_Quantum_Spin_Textures_in_TIs, Valla_et_al_PRL_2011_Spin_ARPES_in_TI_BiSe}.

The mean residual CB population (averaged over a part of the surface Brillouin zone close to the Dirac point), $n_c$, as a function of field amplitude, $F_0$, is shown for the linearly polarized pulse in Fig. \ref{fig:Edited_N}. This population monotonically increases with $F_0$ and has only a weak dependence on the direction of polarization. For a given field amplitude, the maximum CB population is realized for the $y$ polarized pulse, $\theta = \pi/2$, while the minimum is at $\theta \approx \pi/8$ (see inset).

\subsection{Circularly polarized pulse}
\label{CP_Pulse}

\begin{figure}
\begin{center}\includegraphics[width=0.45\textwidth]{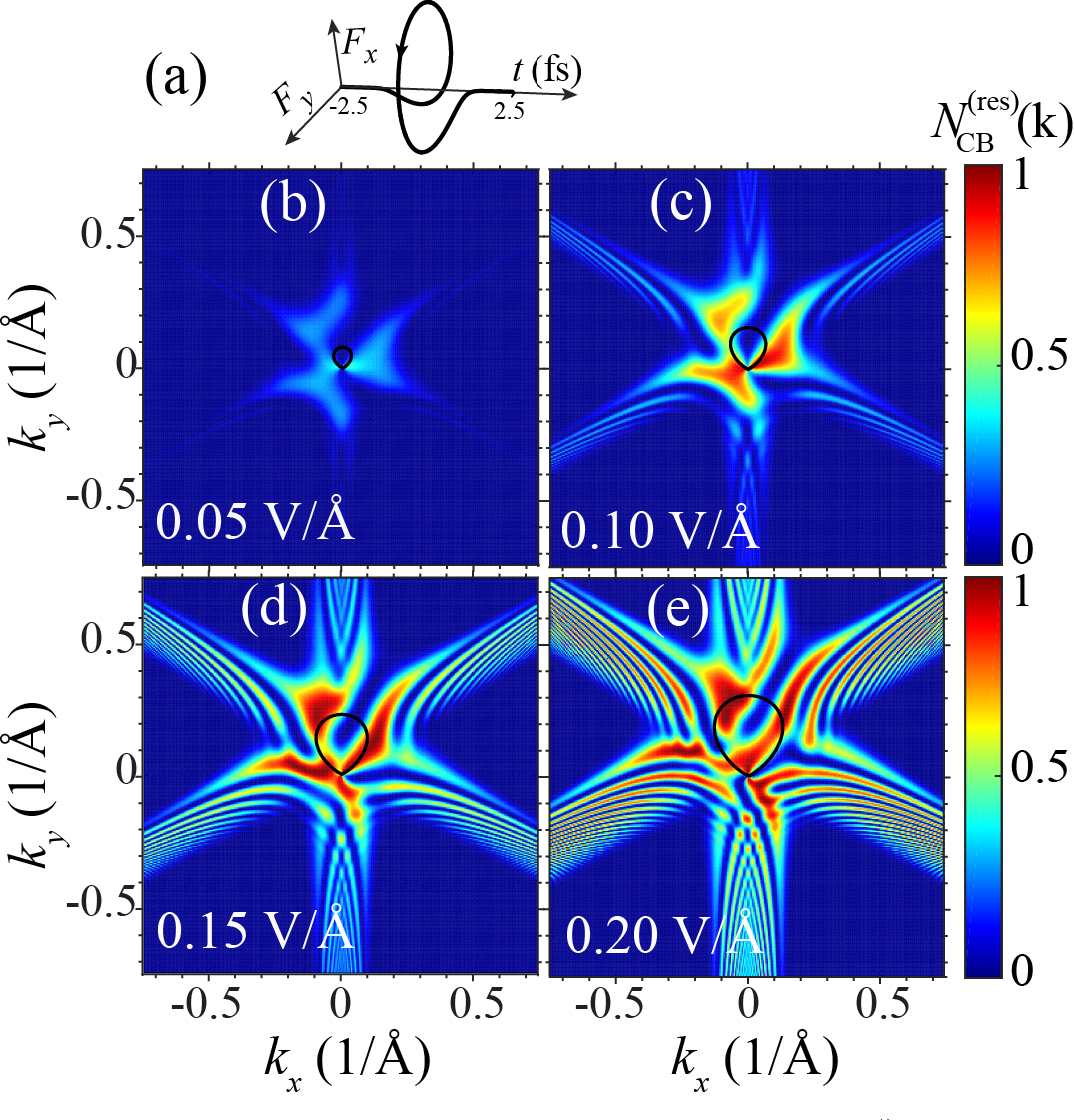}\end{center}
\caption{(color online) Residual CB population as a function of crystal momentum. Excitation pulse is right-handed circularly polarized;  its waveform is shown in panel (a). The amplitude of the pulse is (b) $F_0=0.05$ $\mathrm{V/\AA}$, (c) $F_0=0.10$ $\mathrm{V/\AA}$, (d) $F_0=0.15$ $\mathrm{V/\AA}$, and (e) $F_0=0.20$ $\mathrm{V/\AA}$.  The solid closed black lines display the separatrices (see the text) for the corresponding pulses.
}
  \label{fig:N1_1R}
\end{figure}

\begin{figure}
\begin{center}\includegraphics[width=0.45\textwidth]{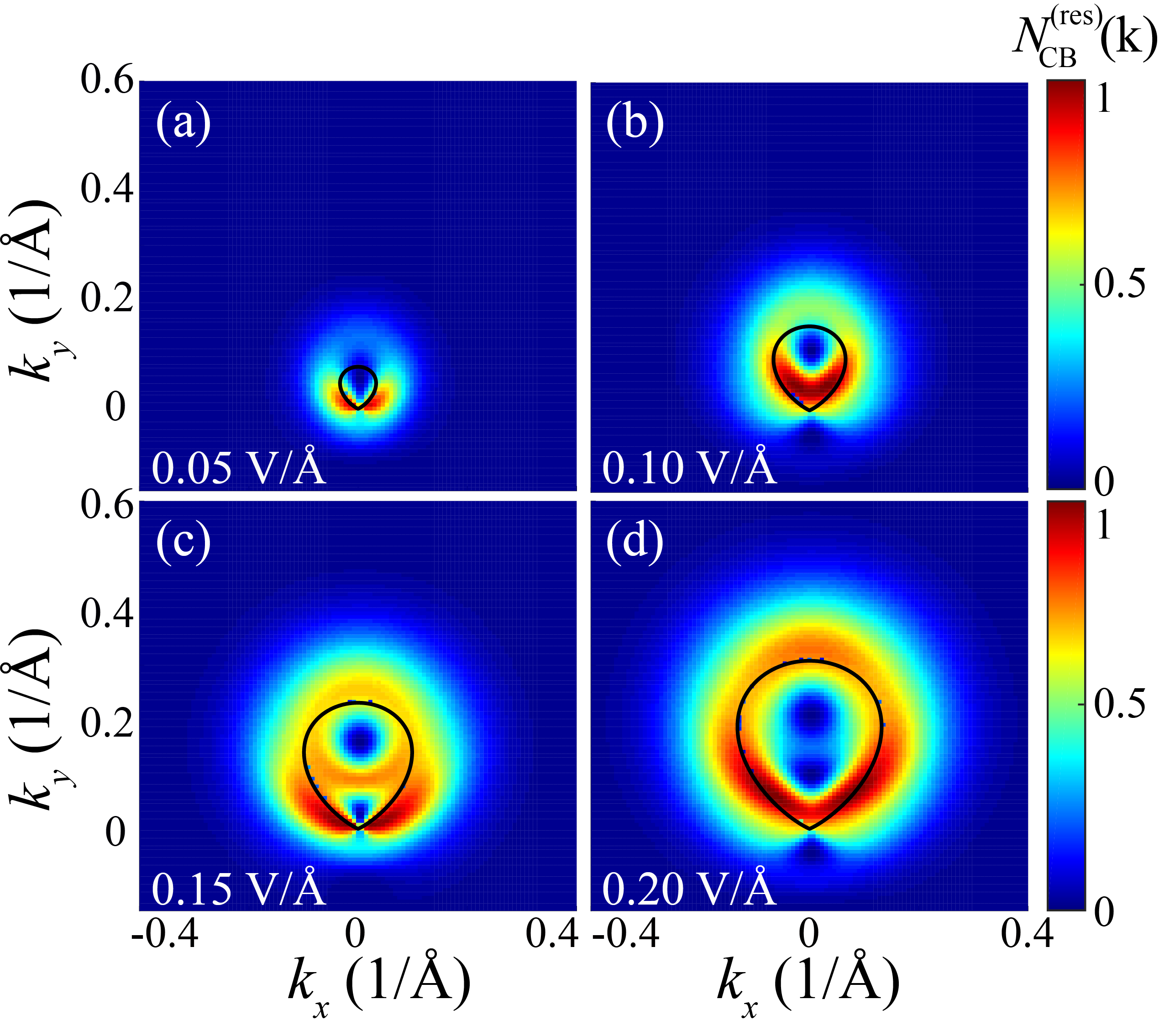}\end{center}
  \caption{(color online) Residual CB population in the vicinity of a Dirac point for graphene for circularly-polarized  left-handed one-cycle pulse with amplitude  of (b) $F_0=0.05$ $\mathrm{V/\AA}$, (c) $F_0=0.10$ $\mathrm{V/\AA}$, (d) $F_0=0.15$ $\mathrm{V/\AA}$, and (e) $F_0=0.20$ $\mathrm{V/\AA}$. The solid closed black lines display the separatrices (see the text) for the corresponding pulses.
 }
  \label{graphene_circular}
\end{figure}

\begin{figure}
\begin{center}\includegraphics[width=0.45\textwidth]{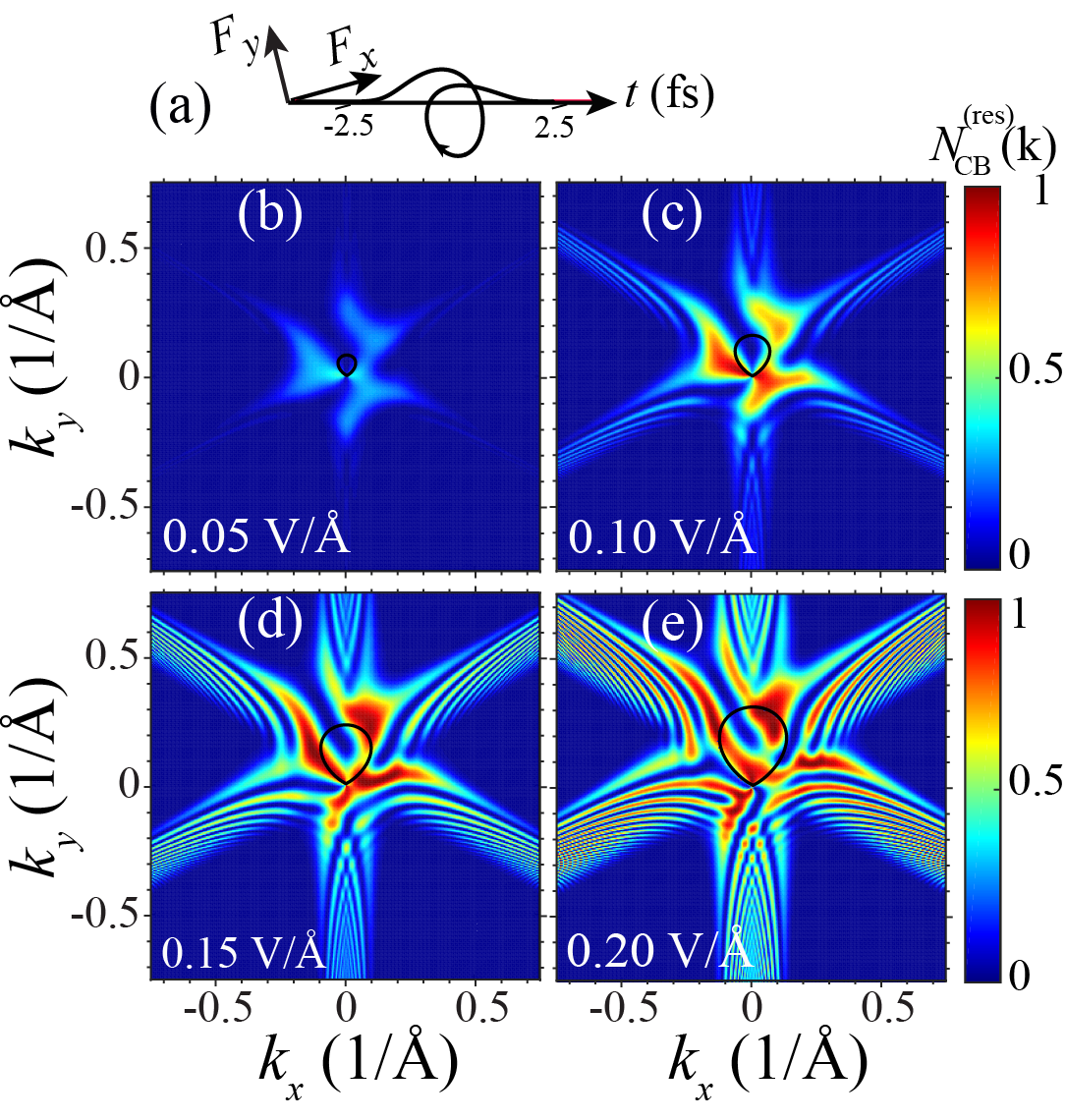}\end{center}
  \caption{(color online) Residual CB population for circularly-polarized  left-handed one-cycle pulse  [waveform is shown on panel (a)] with amplitude  of (b) $F_0=0.05$ $\mathrm{V/\AA}$, (c) $F_0=0.10$ $\mathrm{V/\AA}$, (d) $F_0=0.15$ $\mathrm{V/\AA}$, and (e) $F_0=0.20$ $\mathrm{V/\AA}$. The solid closed black lines display the separatrices (see the text) for the corresponding pulses.
 }
  \label{fig:N1_1L}
\end{figure}

Two-dimensional solids and topological crystals in the reciprocal space possess nontrivial topological properties (chirality) related to the Berry-Zak phase \cite{Berry_Phase_Proc_Royal_Soc_1984, Zak_PhysRevLett.62_1989_Berry_Phase_in_Crystals, Xiao_Niu_RevModPhys.82_2010_Berry_Phase_in_Electronic_Properties}. 
In this Section, we will study this topology using circularly-polarized ultrashort pulses. Note that the linearly-polarized pulses employed in the previous Section are ``blind'' to chirality. 

A one-cycle circularly-polarized pulse we will use has electric field that can be parametrized as the following waveform: 
\begin{eqnarray}
\mathbf F(t)&=&\left(F_x(t),F_y(t) \right),\nonumber\\
 F_x(t)&=& \pm F_0e^{-u^2}(1-2u^2),\nonumber\\
 F_y(t)&=& 2F_0ue^{-u^2},
\end{eqnarray}
where sign $\pm$ determines the handedness of the pulse [$+$ is for a right-hand circularly polarized pulse (RCP), and $-$ is for a left-hand circularly polarized pulse (LCP)]. For brevity, we denote such one-cycle RCP as 1R and one-cycle LCP as 1L. These 1R and 1L pulses are related by reflection in the $yz$ plane ($\mathcal P_{yz}$-reflection). Note that the same change of handedness can be obtained by $\mathcal T$-reversal plus reflection in the coordinate center, i.e., $\mathcal T P_{yz}\mathcal P_{xz}$-transformation. The pairs of pulses with opposite chirality are ideally suited to probe the chirality of the surface Bloch bands because the resulting electron distributions depend on the sign of the Berry curvature with respect to the pulse chirality. 


Following this line, for a 1R pulse, residual CB population distributions in the reciprocal space are shown in Fig. \ref{fig:N1_1R} for several field amplitudes $F_0$. Similar to the linear-polarization case considered above in Sec.\ \ref{Linear_Polarized}, here the residual CB population is also large for $F_0\gtrsim 0.1~\mathrm{V/\AA}$ implying that the electron dynamics is highly irreversible. The CB population distribution shows a chiral pattern, which is correlated with the handedness of the pulse.

A solid closed black curve seen in Fig. \ref{fig:N1_1R} (b)-(e) is the separatrix \cite{Stockman_et_al_PhysRevB.93.155434_Graphene_Circular_Interferometry}. This is defined as a set of initial points $\mathbf q$ for which electron trajectories pass precisely through the Dirac ($\Gamma$) point. Its parametric equation is $\mathbf q(t)= -\mathbf k(0,t)$, where $t\in (-\infty,\infty)$ is a parameter, and $\mathbf k(\mathbf q,t)$ is a Bloch trajectory originating at a point $\mathbf q$ as given by Eq.\ (\ref{kvst}). Thus, the separatrix is an electron trajectory starting at $\mathbf k=0$ (i.e., at the $\Gamma$-point) and reflected in the $xz$ plane ($\mathcal P_{xz}$- reflection).
 For initial crystal momentum $\mathbf q$ inside of the separatrix, the electron trajectory, $\mathbf k(\mathbf q,t)$, encircles the $\Gamma$-point, otherwise it does not. 
 
Because the coupling dipole matrix element (the non-Abelian Berry connection) is large (singular) at the $\Gamma$-point [see Fig.\ \ref{fig:Dipole}], one may expect that  the residual CB population will be enhanced close to the separatrix.  This was the case for graphene in Ref.\ \onlinecite{Stockman_et_al_PhysRevB.93.155434_Graphene_Circular_Interferometry} (see also Fig.\ \ref{graphene_circular} and its discussion below in this Section) but it is not pronounced in the present study, as Fig. \ref{fig:N1_1R} demonstrates. 

A fundamental difference between the TI and graphene is that the non-Abelian Berry curvature, $\bm\Omega_{cv}(\mathbf k)=\frac{\partial}{\partial\mathbf k}\times\mathbfcal A_{cv}(\mathbf k)$, a gauge-invariant field whose integral over reciprocal space area is equal to the Berry phase, for graphene is real and singular -- it has a $\delta$-function-singularity localized at the Dirac ($K$ and $K^\prime$) points. For a circular pulse for graphene, there is only one passage of an electron, which is moving in the reciprocal space, by the Dirac point. Thus there is only one significant amplitude to undergo a VB$\to$CB transition, and, consequently, there is no pronounced interference along the separatrix, see Fig.\ \ref{graphene_circular}. Note that the these illustrative data for graphene were calculated as in Ref.\ \onlinecite{Stockman_et_al_PhysRevB.93.155434_Graphene_Circular_Interferometry}.

In a sharp contrast, in present model of TI, there are regions along the radial lines emanating from the $\Gamma$-point, which are seen as ridges in Fig.\ \ref{fig:Dipole}, where the  non-Abelian Berry connection (dipole matrix element) is increased and possesses a non-trivial phase. These regions overlap in the vicinity of the $\Gamma$-point close to the separatrix. Consequently, the corresponding amplitudes of the VB$\to$CB transitions interfere causing the chiral pattern of interference fringes seen in Fig.\ \ref{fig:N1_1R}. The separatrix itself is surrounded by regions of high CB population but not seen as a continuous arc in contrast to the case of graphene -- cf.\ Fig.\ \ref{graphene_circular} and Ref.\ \onlinecite{Stockman_et_al_PhysRevB.93.155434_Graphene_Circular_Interferometry}. 

For high enough fields, $F_0\gtrsim 0.1~\mathrm{V/\AA}$, there is a pronounced pattern of interference fringes in Fig.\ \ref{fig:N1_1R}(b)-(e). Note that his pattern is a self-referenced electron interferogram whose chirality is due to the nontrivial phase of the non-Abelian Berry connection. This interference is caused by passing the ridge of $\cal A$ twice during one optical cycle and, also, close the separatrix, additionally by passing close to the Dirac point where the VB$\to$CB transitions predominantly occur. The corresponding transition amplitudes interfere generating the fringes, which carry information about both the non-Abelian Berry curvature and the dynamic phase. 



The CB population for the left-handed pulse (1L) is shown in Fig. \ref{fig:N1_1L}. Obviously it is related to the pattern in Fig. \ref{fig:N1_1R} by the $\mathcal P_{yz}$-reflection. This exactly is expected from the symmetry of the problem. Correspondingly, the distributions in Fig. \ref{fig:N1_1L} exhibit opposite chiralities but otherwise are similar to those in Fig. \ref{fig:N1_1R}.



\begin{figure}
\begin{center}\includegraphics[width=0.47\textwidth]{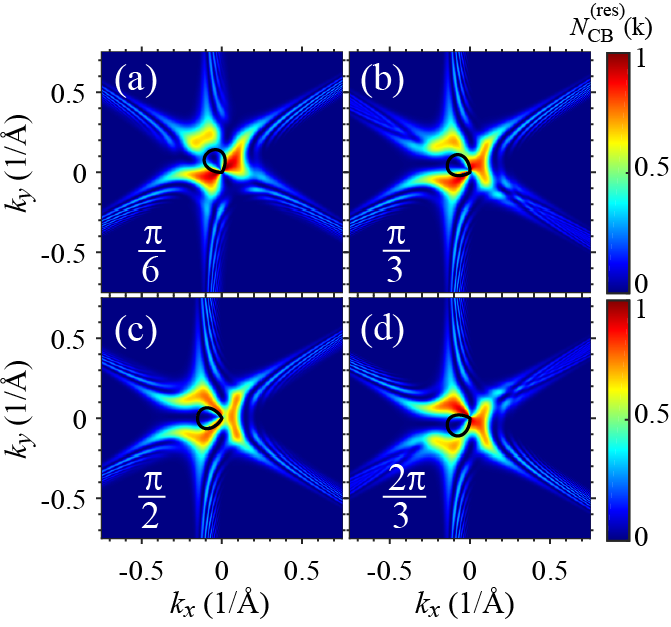}\end{center}
  \caption{(Color online)
  Residual electron population after 1R single-oscillation pulse with amplitude $F_0=0.1$ $\mathrm{ V/\AA}$. The corresponding carrier-envelope phases (i.e., the angle between the maximum field, $\mathrm F_0$) and the positive $x$ axis is indicated in the corresponding panels. 
  }
  \label{one_RC_diff_angle}
\end{figure}

\begin{figure}
\begin{center}\includegraphics[width=0.47\textwidth]{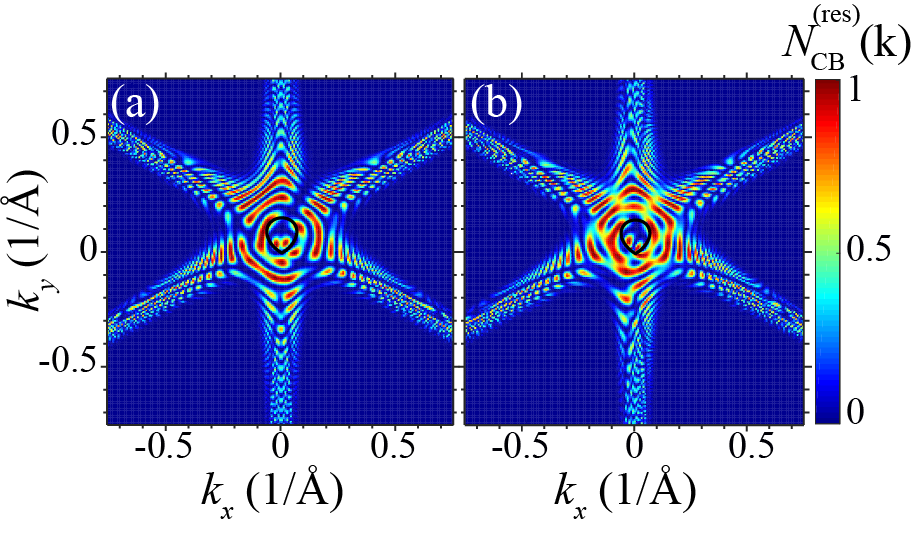}\end{center}
  \caption{(Color online) Residual CB population for two-cycle circularly polarized pulse with the amplitude of  0.1 $\mathrm{V/\AA}$.
The two-cycle pulses are 2R (a) and 1R+1L (b).  The solid closed black lines display the separatrices (see the text).
  }
  \label{fig:RCB_population_1R_1R}
\end{figure}

We consider also pulses consisting of two optical cycles with opposite chiralities whose waveforms are parametrized as  
\begin{eqnarray}
 F_x(t)&=& F_0 \Bigl( e^{-u^2}(1-2u^2)
 \nonumber\\
&& \pm e^{-(u-u_0)^2}(1-2(u-u_0)^2)  \Bigr)~,
 \nonumber\\
 F_y(t)&=&2F_0\left( ue^{-u^2}+(u-u_0)e^{-(u-u_0)^2} \right)~,
\end{eqnarray}
where $u_0=t_0/\tau$ and $\tau=1$ fs. The delay time (the time between the two cycles) is $t_0=4$ fs. Here the plus sign in the 
above expression corresponds to two right-handed cycles (2R), while the minus sign corresponds to one right-handed cycle followed by one
left-handed cycle (1R+1L). Note these opposite-chirality pulses are related by the $\mathcal{P}_{yz}$-reflection instead of the $\mathcal T$-reversal.

The residual CB population for the two-cycle pulse is shown in Fig.\ \ref{fig:RCB_population_1R_1R}(a) for the 2R pulse and Fig.\ \ref{fig:RCB_population_1R_1R}(b) for the
1R+1L pulse. In comparison to the one-cycle pulse, the CB population for two-cycle pulse shows more interference fringes, which is due to the interference of the transition amplitudes accumulated during the two constituent subcycles. 
The CB population distribution also has a chiral pattern for all types of two-cycle pulses (2R and 1R+1L). This chirality is due to non-Abelian Berry phase, which is non-zero due to the warping terms in the Hamiltonian of Eq.\ (\ref{H0}). In a sharp contrast, in graphene the non-Abelian Berry connection is real -- see Eqs.\ (\ref{Axgr})-(\ref{Aygr}). Consequently, the residual CB population for any pulses of the same chirality (1R, 1L, 2R, or 2L) is non-chiral and symmetric with respect to the $\mathcal{P}_{yz}$-reflection \cite{Stockman_et_al_PhysRevB.93.155434_Graphene_Circular_Interferometry}.

As we have already discussed above in conjunction with Eqs.\ (\ref{Dx_TI})-(\ref{Dy_TI}), for the TIs surface reciprocal space, the Berry connection is distributed bringing about a nontrivial phase that leads to chirality of the CB residual electron distribution. In Figs.\   \ref{one_RC_diff_angle} (a)-(d) [see also Fig.\ \ref{fig:N1_1R}(c) for comparison], we illustrate the dependence of this chirality on the orientation of the 1R pulses for a moderate pulse amplitude,  $F_0=0.1~\mathrm{V/\AA}$. Note that this orientation is determined by the angle between the direction of the maximum field, $\mathbf F_0$, and the positive $x$-axis, which is nothing else as the carrier-envelope phase, $\theta$, of the pulse. As we see from these figures, the chirality is present for $\theta=0,\pi/3$, and $2\pi/3$; it is completely absent for $\theta=\pi/6$, and $\pi/3$. Generally, the chirality is absent when the maximum field is in the direction of any of the ridges in Fig.\ \ref{fig:Dipole}(a), i.e., $\theta=\frac{\pi}{6}+n\frac{\pi}{3}$, where $n$ is an integer; in contrast, it is maximum when the maximum field is along the bisector between the ridges, $\varphi=n\frac{\pi}{3}$. Note that in all cases of Fig.\ \ref{fig:N1_1R}(c) and Figs.\  \ref{one_RC_diff_angle}(a)-(d) , the electron distribution consists of ``bright'' regions of a high population adjacent to and outside of the separatrix. These bright regions have peripheral ``wings'' extending along the six ridges of the non-Abelian Berry connection [Fig. \ref{fig:Dipole}(a)] $\mathbfcal A_\mathrm{cv}(\mathbf k)$.

\section{Conclusion Discussion}

Electron dynamics on the surface of $\mathrm{Bi_2Se_3}$ in the field of an ultrashort and strong optical pulse results in a significant CB population during the pulse and after the pulse. This residual CB population is large and comparable to the maximum CB population during the pulse, which implies that the electron dynamics is highly irreversible. For a linearly polarized pulse, the electron dynamics significantly depends on the polarization direction of the pulse. 
There is a pattern of interference fringes in electron CB distribution [see Figs.\ \ref{fig:RCB_TI} and \ref{fig:RCB_graphene}]. These fringes appear as a result of interference of two events of the electron passage by the Dirac ($\Gamma$) point during the single optical oscillation. These are somewhat similar to those predicted earlier for graphene \cite{Stockman_et_al_PRB_2016_Graphene_in_Ultrafast_Field} (the present field is relatively low and only one pair of fringes appear). Note that limited preliminary results on linear-polarized pulses interacting with TI's, which included only Fig.\ \ref{fig:RCB_TI} of the present article, were recently published in Conference Proceedings \cite{Oliaei_Motlagh_et_al_2017_TI}.

Circularly-polarized (chiral) single-oscillation pulse induces a chiral distribution of the residual CB population in the reciprocal space. The handedness of the induced chirality is determined by the handedness (right or left) of the pulse.  Such a chiral response in $\mathrm{Bi_2Se_3}$ is due to the warping terms in the low-energy surface Hamiltonian near the Dirac point, which is the $\Gamma$-point in TI's. This leads to the non-Abelian Berry connection being complex with phase winding that causes the aforementioned chirality. In a sharp contrast, in graphene the non-Abelian Berry connection is real, and no chirality is induced by a single-oscillation chiral pulse.

The electron interference patterns predicted in this work are self-referenced electron holograms that carry rich information about topological properties (the non-Abelian Berry connection and curvature) of the TI reciprocal space. Such self-referenced holograms can be measured using time-resolved angle-resolved photoelectron spectroscopy (TR-ARPES) \cite{Chiang_et_al_PhysRevLett.107_Berry_Phase_in_Graphene_ARPES, Gierz_Snapshots-non-equilibrium-Dirac_Nat-Material_2013}. In principle, it may be possible to restore the topology of the Bloch bands from these holograms. We will consider the latter problem elsewhere. 

Optical pulses, both linearly and circularly polarized, create generally asymmetric carrier distributions in the reciprocal space. These will manifest themselves as currents (net charge transfer) in the real space -- cf.\ Refs.\ \onlinecite{Schiffrin_at_al_Nature_2012_Current_in_Dielectric, Stockman_et_al_Sci_Rep_2016_Semimetallization, Higuchi_Hommelhoff_et_al_Nature_2017_Currents_in_Graphene}. The corresponding charge transfer per pulse can be measured macroscopically. For linearly polarized pulses, the direction of the net charge transfer is parallel to the maximum electric field. The net charge transferred (sign and magnitude) is defined by the carrier-envelope phase of the pulse and provides a direct access to its measurement \cite{Stockman_et_al_Nat_Phot_2013_CEP_Detector}.

For circularly polarized pulses considered in this article, the resulting electron distributions in the reciprocal space are generally both asymmetric and chiral. Correspondingly, there are two types of currents: (i) Direct current in a direction parallel to the maximum field and (ii) Hall current in a normal direction. The latter depends on the chirality of the pulse vs. the chirality of the Bloch bands at the $\Gamma$-point. There is a circular current present during the pulse that follows the shape of the TMDC specimen. This current will produce THz radiation as in Ref.\ \onlinecite{Huber_et_al_s41586-018-0013-6_Nature_2018_Valleytronics}, which can also provide an access to the ultrafast electron dynamics. We will publish our results on the currents induced in TI's elsewhere.

\begin{acknowledgments}
This work by MIS was supported by Grant DE-FG02-01ER15213 from the Atomic, Molecular and Optical Sciences Program, Office of Basic Energy Sciences, the US Department of Energy. Work of VA was supported by a grant DE-SC0007043 from the Physical Behavior of Materials Program, Office of Basic Energy Sciences, and the US Department of Energy. SAOM gratefully acknowledges support by MURI Grant FA9550-15-1-0037 from the US Air Force Office of Scientific Research. JSW support came from EFRI NewLAW Grant EFMA-17 41691 from US National Science Foundation.
\end{acknowledgments}

%

\end{document}